**Advanced scanning probe lithography**

**Ricardo Garcia[1], Armin W. Knoll[2] and Elisa Riedo[3]**

1. Instituto de Ciencia de Materiales de Madrid, CSIC, Sor Juana Inés de la Cruz 3. 28049 Madrid, Spain

2. IBM Research - Zurich, Saeumerstr. 4, 8803 Rueschlikon, Switzerland

3. School of Physics, Georgia Institute of Technology, Atlanta, GA, USA

**The nanoscale control afforded by scanning probe microscopes has prompted the development of a wide variety of scanning probe-based patterning methods. Some of these methods have demonstrated a high degree of robustness and patterning capabilities that are unmatched by other lithographic techniques. However, the limited throughput of scanning probe lithography has prevented their exploitation in technological applications. Here, we review the fundamentals of scanning probe lithography and its use in materials science and nanotechnology. We focus on the methods and processes that offer genuinely lithography capabilities such as those based on thermal effects, chemical reactions and voltage-induced processes.**

Progress in nanotechnology depends on the capability to fabricate, position, and interconnect nanometre-scale structures. A variety of materials and systems such as nanoparticles, nanowires, two-dimensional materials like graphene and transition metal dichalcogenides, plasmonics materials, conjugated polymers and organic semiconductors are finding applications in nanoelectronics, nanophotonics, organic electronics and biomedical applications. The success of many of the above applications relies on the existence of suitable nanolithography approaches. However, patterning materials with nanoscale features aimed at improving integration and device performance poses several challenges. The limitations of conventional lithography techniques related to resolution, operational costs and lack of flexibility to pattern organic and novel materials have motivated the development of unconventional fabrication methods[1-3].

Since the first patterning experiments performed with a scanning probe microscope in the late 80s, scanning probe lithography (SPL) has emerged as an alternative lithography for academic research that combines nanoscale feature-size, relatively low technological requirements and the ability to handle soft matter from small organic molecules to proteins and polymers. Scanning probe lithography experiments have provided striking examples of its capabilities such as the ability to pattern 3D structures with nanoscale features[4], the fabrication of the smallest field-effect transistor[5] or the patterning of proteins with 10 nm feature size[6].

Figure 1a shows a general scheme of SPL operation. There is a variety of approaches to modify a material in a probe-surface interface which have generated several SPL methods. Scanning probe lithographies can be either classified by emphasizing the distinction between the general nature of the process, chemical versus physical, or by considering if SPL implies the removal or addition of material. However, we consider it is more inclusive and systematic to classify the different SPL methods in terms of the driving mechanisms used in the patterning process, namely thermal, electrical, mechanical and diffusive methods (Fig. 1b).

**Challenges in nanoscale lithography**

The workhorse of large volume CMOS fabrication, optical lithography at a wavelength of 192 nm, has reached the physical limits in terms of minimal achievable pitch of a single patterning run of about 80 nm. To make denser integrated circuits and additionally to keep Moore's law fulfilled for feature sizes approaching the single digit nm range, multi-patterning extensions such as double and triple-patterning have been introduced with the acceptance of the inherently higher manufacturing costs. Alternatively, the technically challenging switch to shorter wavelengths in extreme UV (EUV) at 13.5 nm is considered a viable although extremely costly possibility for the next years[7].

Economic reasons dictate throughputs of more than 100 wafers/h corresponding to >$10^{12}$ $\mu m^2$/h for high volume production techniques (Fig. 2a). This mask-based high-volume lithography environment needs of accompanying techniques for flexible low-volume production, mask fabrication and prototyping of the next generation devices.

These applications require flexible tools without the overhead to produce masks for each patterning step, so called mask-less lithography technologies. Their throughput scales phenemenologically with the achievable resolution according to a power law as it has been first recognized by Tennant[8]. Among the mask-less methods the dominating technique is electron beam lithography (EBL), using a Gaussian electron beam (GEB) or variable shaped beams (VSB) for sub-20 nm resolution or high throughput demands, respectively. At high resolutions the trade-off between resolution and throughput is determined by the sustainable beam current and resist sensitivity. Both higher currents and enhanced sensitivity through chemical amplification (chemically amplified resists, CAR) lead to a reduction in resolution, limiting the throughput at high resolutions (GEB). Even higher resolutions in the single digit nm range can be obtained by using inorganic resists or electron beam induced deposition (EBID)[9], albeit at very limited throughputs of about 1 µm$^2$/h and high costs. Currently massively parallel approaches are under study with the goal to scale EBL towards high volume production [10,11]. At the same time alternative nano-patterning methods have been explored. Novel beam based methods using He[12] and Ne[13] ions instead of electrons promise high resolution and enhanced resist sensitivity.

In parallel to the developments of beam-based methods, scanning probe lithography (SPL) methods are receiving renewed interest because of their flexibility to handle novel materials and their inherent inspection and positioning capabilities. Since their invention, scanning probe microscopes have been used to image, modify and manipulate surfaces at the nanometer and atomic scales. Atomic scale manipulations have been performed in ultra-high vacuum although the exceedingly small throughput values greatly limit their impact and applications. Recent developments for ambient atmosphere have shown that some scanning probe nanolithography approaches could also be competitive in terms of resolution, throughput and versatility of the materials that can be patterned. This makes SPL an appealing nanolithography for research and some niche technological applications. For example, thermal SPL has achieved a resolution of 10 nm while the throughput is in the $10^4$-$10^5$ µm$^2$/h range.

**The role of Scanning Probe Lithography**

Scanning probe lithography includes several approaches to pattern materials with nanoscale resolution (Fig. 1b). These approaches have a common thread, which is the use of a scanning sharp probe to produce local modifications on a surface. The variety of SPL approaches arises from two main factors. 1) The wealth of processes that could be controlled by using a sharp probe in contact or near contact with a nanoscale region of a sample surface. The processes involved in SPL imply mechanical, thermal, electrostatic and chemical interactions, or different combinations among them. 2) The various methods to control the position of the scanning probe relative to the underneath surface, for example through quantum tunnelling between the probe and a conductive surface as in the scanning tunnelling microscope (STM), or by controlling the force between the probe and the surface as in the standard atomic force microscope (AFM). In fact, most of the current SPL methods rely on the use of an AFM.

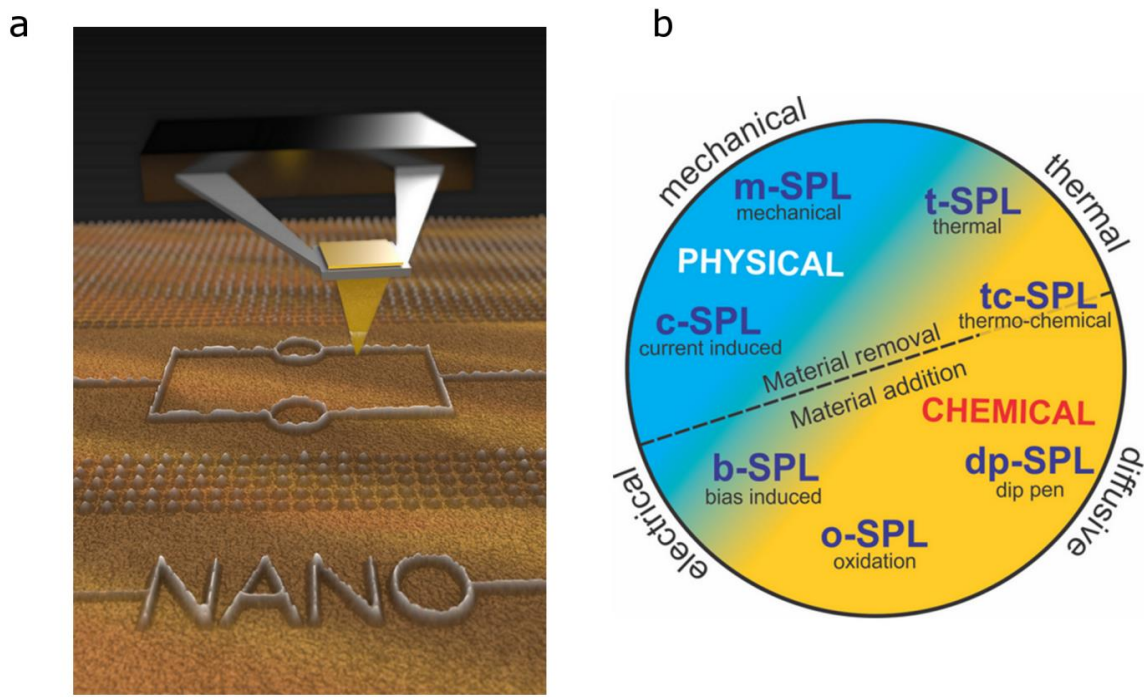

**Figure 1. Scanning probe lithography**. **a**. Schematic of SPL where imaging and patterning applications are orthogonal. **b**. Classification of SPL methods according to the dominant tip-surface interaction used for patterning, namely electrical, thermal, mechanical and diffusive processes.

The potential and variety of the methods available to scanning probe microscopy (SPM) to locally modify surfaces was already evident in the early experiments. However, many of those approaches albeit inspiring by their atomic scale manipulation capabilities have been proved unpractical for any large scale patterning or device applications. In this Review the focus is on the SPL methods that are robust and flexible enough to make patterns and/or devices with a high degree of reproducibility and show a potential for scalability and compatibility with ambient conditions and novel materials. These methods are collectively called advanced scanning probe lithography.

Compared to other techniques such as EBL, the principal advantage of SPL is that it is a single step process with sub-10 nm resolution. Most of the SPL writing processes are 'direct write' in nature, creating structures on the fly without the need of a resist or a subsequent development step. This is in particular relevant for patterning novel types of functional materials such as graphene or other 2D materials, which are known for being sensitive to resist residuals[14]. Most SPL methods operate under (controlled) atmospheric conditions, which reduces the tool overhead and costs. It also facilitates its applications. The constituents of the atmosphere may even provide the functionality for some SPL methods such as bias SPL or oxidation SPL. The simplicity of the techniques also allows for straightforward parallelization schemes. Furthermore, the scanning probe microscope is capable of detecting surface features down to atomic resolution. In contrast to beam-based methods, imaging and patterning in SPL are orthogonal, i.e., the imaging process does neither influence the written structures nor implies a partial writing operation. Together, the non-destructive imaging capability and the direct writing enables to establish the concept of so called 'closed loop lithography', i.e., a lithography tool with inherent feedback of the writing result to optimize the writing stimuli on the fly. This tool is thus capable to autonomously control the writing process, improving dramatically the ease of use to create complex and high resolution nanoscale structures.

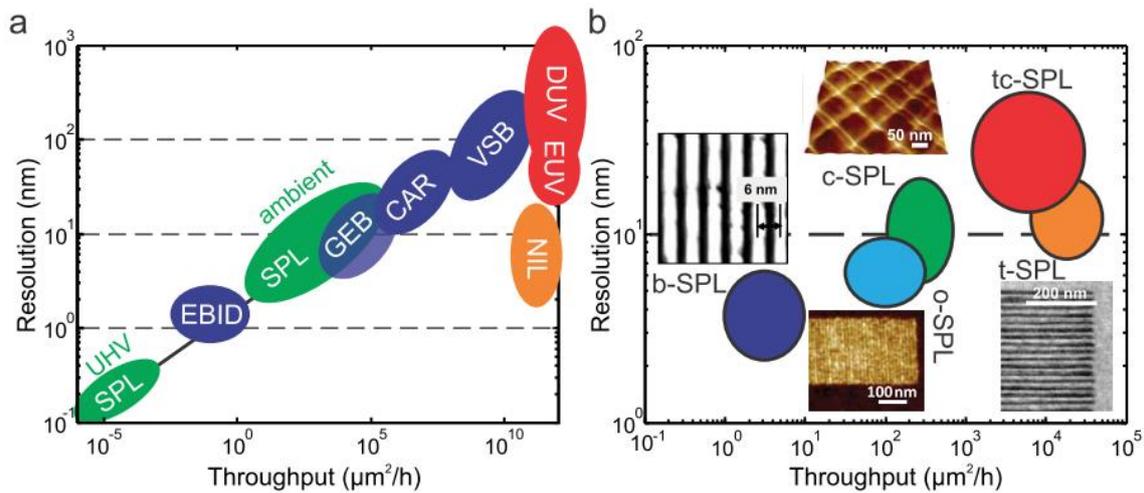

**Figure 2. Nanofabrication landscape. a.** Resolution and throughput in nanolithography. High volume techniques require throughput values $> 10^{12}$ μm$^2$/h. At lower throughput, mask-less electron beam and scanning probe techniques converge roughly on a single line, called Tennant's law. NIL stands for nano imprint lithography. **b**. Within the advanced scanning probe techniques a similar correlation exists. High resolution results are shown for bias SPL (b-SPL), oxidation SPL (o-SPL), current controlled SPL (c-SPL), thermo-chemical SPL (tc-SPL) and thermal SPL (t-SPL). Figure reprinted with permission from: a, ref. 8, © Springer; b; images for b-SPL, ref. 61, © American Chemical Society; c-SPL, ref. 51, © SPIE; t-SPL, ref. 27, © American Chemical Society.

In general, the ability of SPL to image the surface of a material and *in-situ* fabricate complex patterns with sub-10 nm precision in size and single nm accuracy in positioning, as well as post-patterning *in-situ* metrology is rather unique. Finally, SPL is compatible with patterning a large variety of materials, including polymers and biological matter. Applications of SPL to pattern silicon[15], graphene[16], piezoelectric/ferroelectric ceramics[17], polymers[4,18-21], proteins[6] have been demonstrated. Very appealing is also the capability to use the same SPL setup to pattern different materials at the same time[22].

**Thermal and Thermochemical SPL**

Thermal scanning probe lithography (t-SPL) was first developed for data storage purposes in the early 90s[23]. In that work it was understood that the transport of heat is only significant if the tip and the sample are in intimate contact. Thus, the heat is highly localized at the tip-sample contact area, which is of the order of a few nm$^2$ due to the nanoscale dimensions of the SPM tip. Furthermore, similar to the case of light, manipulation by heat does not require the presence of conductive surfaces and thus is widely applicable.

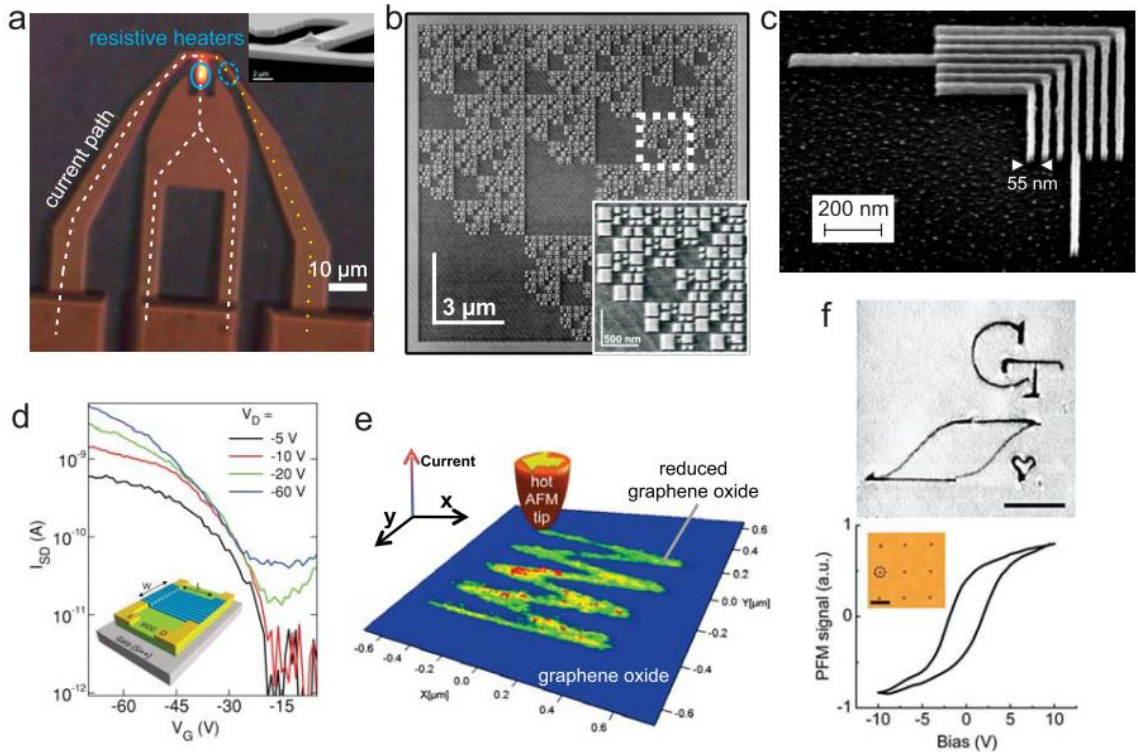

**Figure 3. Thermal and Thermochemical SPL. a.** Silicon thermal cantilever comprising integrated joule heaters for tip heating and for thermal sensing. **b.** High speed t-SPL. Topographical image of a fractal pattern comprising 880 x 880 pixels written in 12.8 seconds.[29] **c.** Silicon structures created from a RIE etch transfer of t-SPL written nested L-lines at 27 nm half-pitch.[27] **d.** Direct patterning of field effect transistors by conversion of a precursor material into pentacene.[33] **e.** Direct thermal conversion of graphene oxide to conductive graphene using tc-SPL[34]. **f.** Top: local crystallization by tc-SPL of a precursor film on plastic or Si to form nanostructures of PbTiO$_3$ (PTO) ceramics. Bottom: Piezo force microscopy measurement of the typical ferroelectric hysteresis loop acquired on a PTO nanodot fabricated by tc-SPL, and shown in the inset[17]. Scale bars 1 μm. Figures reprinted with permissions from: a, ref. 126, © SPIE; b, ref. 29, © Institute of Physics; c, ref. 27, © American Chemical Society; d, ref 31, © Wiley; e, ref 32, © American Association for the Advancement of Science; f, ref. 17, © Wiley.

Heat is used in thermal and thermochemical SPL to modify mechanically or chemically a material. In the early experiments, laser heating with pulse times of microseconds and linear scan speeds of 25 mm/s demonstrated the high speed potential of thermo-mechanical writing schemes. Today, heaters integrated into silicon SPM cantilevers are used (Fig. 3a). The integration improves the resolution and facilitates the control of the writing parameters. The tip is resistively heated by a current flowing in the cantilever legs, which are highly doped except for the region where the tip is positioned. In silicon the maximum sustainable temperature at the heater position is limited by electro-migration of the dopants to 800-1000°C, depending on the type of dopant. Typical thermal time constants of the integrated heaters range from 5 to >100 μs [24], allowing for fast switching of the thermal stimulus. The effective temperature at the substrate surface depends on the ratio of thermal resistance of the substrate and of the combined resistance of the tip and tip-sample interface[24]. For sharp tips with radii on the order of 5 nm, and polymer films thicker than the lateral size of the contact, the temperature of the heater is reduced by about a factor of two at the polymer surface. Thus, highly temperature sensitive materials are required at high resolution. We also note that, in ambient conditions, the thermal heater may also act as a height sensor and imaging can be achieved by using only electrical control without the need for an optical-lever setup.

Figure 3 summarizes some recent achievements in thermal scanning probe lithography. In all cases presented in this figure the highly localized heat stimulus is used to trigger a nanoscale reaction, which consists of excitation or cleavage of physical or chemical bonds, as well as more complex reactions such as crystallization processes. We distinguish the thermal patterning methods according to the characteristics of the created patterns. If the thermal process results in efficient removal of material for the purpose of generating a topographical pattern, the method is termed thermal SPL (t-SPL). If the process is purely thermochemical in nature[25] and the resulting patterns are made of a material with structure and chemistry different from the original one, we term the method thermochemical SPL (tc-SPL), also known as thermochemical nanolithography (TCNL). In t-SPL[26,27] either molecular glass resists[4] or the thermally responsive polymer polyphthalaldehyde (PPA)[28] are used as substrate, and they perform exceptionally well for topographic patterning. In PPA the fission of a single bond is amplified by spontaneous decomposition of the remaining polymer chains resulting in a highly efficient patterning process. The patterns shown in Fig. 3b contain 880 x 880 pixels and were written in less than 12 s, demonstrating the high throughput of the approach.[29] Throughputs are in the range of 5 x $10^4$ μm$^2$/h ( Fig. 2b). Patterning at a half pitch down to 10 nm without proximity corrections was demonstrated[27]. Other milestones towards technical readiness of the technique are the stitching of patterning fields at < 10 nm precision[30] and a high-quality pattern transfer into the underlying silicon substrate at high resolution and low line edge roughness (Fig. 3c).[27]

A key example of thermochemical SPL is the use of hot probes for on demand patterning of field effect transistors from a pentacene precursor as shown in Fig. 3d.[33] Thermal reduction of functionalized graphene by tc-SPL is also perceived as an attractive way to pattern graphene with nanoscale precision. Reduced graphene oxide[34] (Fig. 3e) and reduced graphene fluoride[35] nanoribbons have been fabricated with width as low as 12 nm and tunable conductivity over four orders of magnitude. Field effect transistors have also been demonstrated by using these nanowires. In addition, direct writing of ferroelectric/piezoelectric ceramic nanostructures on plastic, glass and silicon was demonstrated by local tc-SPL induced cristallization[17] (Fig. 3f). Recently, tc-SPL was also used to deprotect active groups such as carboxylic acid[25,31] and amine groups[32], which can subsequently be used for biochemical conjugation of nano-objects. Multifunctional patterns of proteins, DNA and $C_{60}$ have been obtained with a resolution down to 10 nm at patterning speeds up to mm/s.[32]

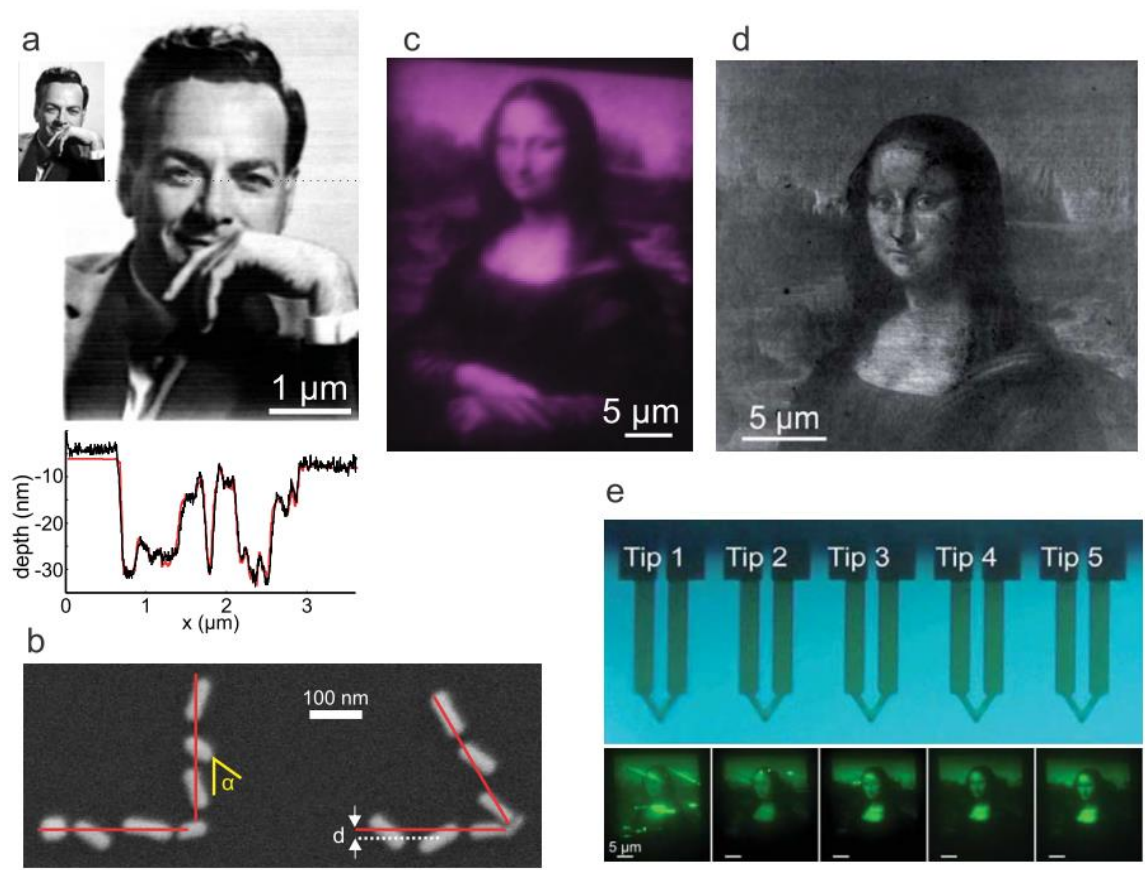

**Figure 4. Grayscale t-SPL and tc-SPL. a.** Grayscale patterning of a photograph of Richard Feynman (Courtesy of the Archives, California Institute of Technology; used with the permission of Melanie Jackson Agency, LLC) with single nanometer absolute depth precision written by closed loop t-SPL. Inset: Programmed bitmap, bottom: cross-sectional profile according to the dotted line. **b.** Precise positioning of Au nanorods on silicon wafer after template removal.[36] The red lines mark the position of the guiding structures. A precision d of 10 nm was achieved (1 sigma). **c.** tc-SPL is used to control the density of amine groups on a polymer film. The thermally deprotected amines are then labelled with a fluorescent dye for visualization, showing in pink the optical fluorescence image of a Mona Lisa picture[20]. **d.** AFM topography image (full z-range 20 nm) of a 3D Mona Lisa image nanopatterned by tc-SPL conversion of a precursor film into PPV [22]. **e.** Image of an array of five thermal cantilevers and corresponding five fluorescence images of five PPV Mona Lisa patterns obtained with that array [22]. Figures reprinted with permissions from: b, ref 36 © American Chemical Society, c, ref. 20 American Chemical Society; d,e: ref 22 © Royal Chemical Society.

**Grayscale chemical and topographical patterning.** A particular strength of t-SPL and tc-SPL is the precise control of the patterning parameters at the nanometer size scale and microsecond time scale. The control is due to the highly reproducible motion of the cantilever and thus the repeatability of the interaction times and forces, and the stability of the heater temperature. In the lithography application it enables the production of three dimensional grayscale relief patterns[28] even on rough surfaces as well as chemical gradients[20] in a single patterning run. By exploiting the 'closed loop lithography' scheme mentioned above, the absolute patterning depth in a PPA polymer film can be controlled to about single nanometer precision, less than the linear dimension of a single resist molecule. An example is shown in Figure 4a depicting the final topographical pattern imaged during the closed loop writing process. Using the same colour scale the inset shows the programmed bitmap, a portrait of R. Feynman who in 1959 wrote a visionary essay about the possibilities of manipulating matter at the micro, nano and atomic scale. The precision achieved can be seen from the cross sectional profiles of both bitmaps at the bottom of Figure 4a. The three dimensional shape of the topographical relief structures was exploited for a precise and oriented positioning of Au nanorods into t-SPL defined guiding structures[36]. Figure 4b shows a SEM image of bare Au nanorods placed with an accuracy of 10 nm (1 sigma) on the silicon substrate after removal of the polymer template containing the guiding structures. Another application of the precise depth control is multilevel data storage, encoding three bit levels into the depth of the indents[37]. A bit error rate of $10^{-3}$ could be achieved.

A similar control was obtained for the degree of chemical functionalization in a thermally sensitive polymer (Fig. 4c) [20]. The heated tip deprotects a functional group in the polymer to unmask primary amines, which serve as attachment sites for subsequent selective functionalization with the desired species of molecules and nano-objects. The density of amine groups on the surface is precisely controlled by the applied temperature and scanning speed, and can be predicted by using an Arrhenius model for the thermally activated chemical reaction. To visualize the programmed gradient of amines on the polymer surface the deprotected functional amine groups are fluorescently labelled using a *N*-Hydroxysuccinimide (NHS) fluorescent dye and imaged with fluorescence microscopy. In Fig. 4c, we show the resulting optical image of a fluorescent "Monna Lisa". Other examples are the conversion of precursors into semiconducting polymers, thus enabling the direct 3D writing of polymers relevant for organic electronics devices (Fig. 4d and 4e). This was first demonstrated by direct patterning of fluorescent structures from a PPV precursor material [18,19,22].

**Bias-induced SPL**

Force microscopy offers a flexible and versatile interface to control chemical processes at the nanoscale. The small size of the AFM tip's apex and the proximity of the surface facilitates the generation of extremely high electrical fields and, in conducting samples, a focused electron current. Remarkably, high electric fields ~10 V/nm (10 GV/m) can be achieved by applying moderate voltages (~10 V). Those fields and/or the associated electron currents are used to confine a variety of chemical reactions and/or to decompose gas[38,39] or liquid[40] molecules that lead to either a locally controlled deposition or to the growth of material on a surface. In addition, b-SPL experiments can be performed in ambient or liquid environments which in turn increases the number of available chemical species.

There is a large variety of methods that combine the application of a voltage with a tip-surface interface to produce nanoscale features. The electric bias across the tip-surface interface can induce local and bulk electrochemical processes such as the anodic oxidation of semiconductors[41,42] and metals[42], the reduction of earth metal oxides[43], metal salts[44] and ionic conductors[45]. Experimental schemes that are closer to the conventional electrochemical set-ups with reference, counter and working electrodes are also being used to deposit metal nanostructures[46,47]. In this context, b-SPL has been used to locally catalyse the reduction of insulating graphene oxide in the presence of hydrogen. Nanoribbons with widths ranging from 20 to 80 nm and conductivities of $>10^4$ $Sm^{-1}$ have been successfully generated, and a field effect transistor was produced[48]. The method involves mild operating conditions, atmospheric pressure and low temperatures (≤115°C). Oxidation SPL, the most robust and established nanolithography method of this kind, is described in the next section.

Bias-induced SPL can involve other processes such as field-induced deposition of matter[39,49-50] current-induced transformations[51,52] and desorption processes[42], the direct deposition of charges[54] or the inversion of the polarization of a local volume in a ferroelectric film[55,56]. The atomic scale resolution potential of b-SPL is illustrated by experiments reporting the local electron –induced hydrogen desorption on a Si(100) surface[53]. In this way clean and H-passivated regions on the surface have been produced. The chemical contrast between those regions has been combined to fabricate the smallest lithographically engineered electron devices[5,57-58]

The electric field at the tip-surface interface can invert the polarization of a small region in a ferroelectric film. This generates a nonvolatile ferroelectric domain. This mechanism has been proposed for data storage[55,59-60]. The nondestructive write-erase process has achieved areal densities of 3.6 Tbit/$inch^2$.

Bias SPL has also produced some other milestones such as the fabrication of the smallest pattern made at ambient pressure and room temperature on a silicon surface[61]. It has also been applied to integrate dissimilar materials with nanoscale accuracy such as Ge patterns on a silicon surfaces[62]. The potential of b-SPL goes beyond the field of nanolithography. The method has been applied to understand new processes to decompose very stable chemical species such as carbon dioxide[39]. These results expand and strengthen the applications of SPL and nanochemistry[45].

**Oxidation SPL**

The discovery of probe-based oxidation[41] was shadowed by the more exciting experiments reporting either modifications[62] or manipulations[63] of surfaces with atomic-scale capabilities[64]. However, the generality and robustness of the underlying chemical process (anodic oxidation) has transformed Dagata's observation into a reliable a versatile nanolithography approach for patterning and device fabrication[42]. Nanopatterning examples range from the generation of arrays of sub-micrometre lines and dots on crystalline surfaces[66-68], self-assembled monolayers[69-71], polymers[72] to the fabrication of nanoscale templates for the growth of single molecule magnets[73], proteins[6], nanoparticles[74-76] to the directed self-assembly of block co-polymers[77], polymer brush nanostructures[78], carbon nanotubes[79] or semiconductor nanostructures[80]. Examples of nanoscale devices and prototypes include, among others, single photon detectors[81], photonic nanocavities[82], quantum devices such as quantum point contacts[83-84], dots[85,86], rings[87] and several graphene devices[89-92]. A variety of transistors such as single electron[93], metal-oxide[94] or nanowire field effect transistors[15,95] have been fabricated by o-SPL. Other applications include the use of the local oxide as a coating to embed nanoparticles on a silicon surface[96]. Oxidation SPL is also contributing to bring a renewed interest to the physics of water meniscus[97-98]. Oxidation SPL has received various names such as local oxidation nanolithography, scanning probe oxidation, nano-oxidation or local anodic oxidation.

The widespread academic use of o-SPL is explained by three features. First, the ability to nanopattern a wide variety of materials ranging from metals to semiconductors to self-assembled monolayers and more recently to graphene or polymer-based resists. Second, its minimal technological requirements which in combination with operation at room temperature and atmospheric pressure conditions makes o-SPL very attractive for academic research. Third, the method has multitask capabilities because at the same time it generates a thin dielectric, a mask for further etching or a template to direct the growth of molecular architectures.

Oxidation SPL is based on the spatial confinement of an anodic oxidation reaction between the tip and the sample surface. The oxidation process is mediated by the formation of a nanoscale water bridge (Fig. 4a)[99]. In fact, due to the sequential character of the o-SPL the generation of a nanopattern might involve the formation of multiple water bridges. The role of the water meniscus is two-fold. It acts as a nanoscale electrochemical cell that provides the oxyanions by which the reaction takes place. In addition, it confines the reaction laterally, this is, the size of the meniscus determines the resolution of the features obtained by this technique[99]. The polarity of the voltage is in such a way that the tip acts as the cathode (negative) and the sample surface is the anode (positive) (Fig. 4c). Oxidation SPL can be either performed with the tip in contact with the sample surface or in a non-contact mode.

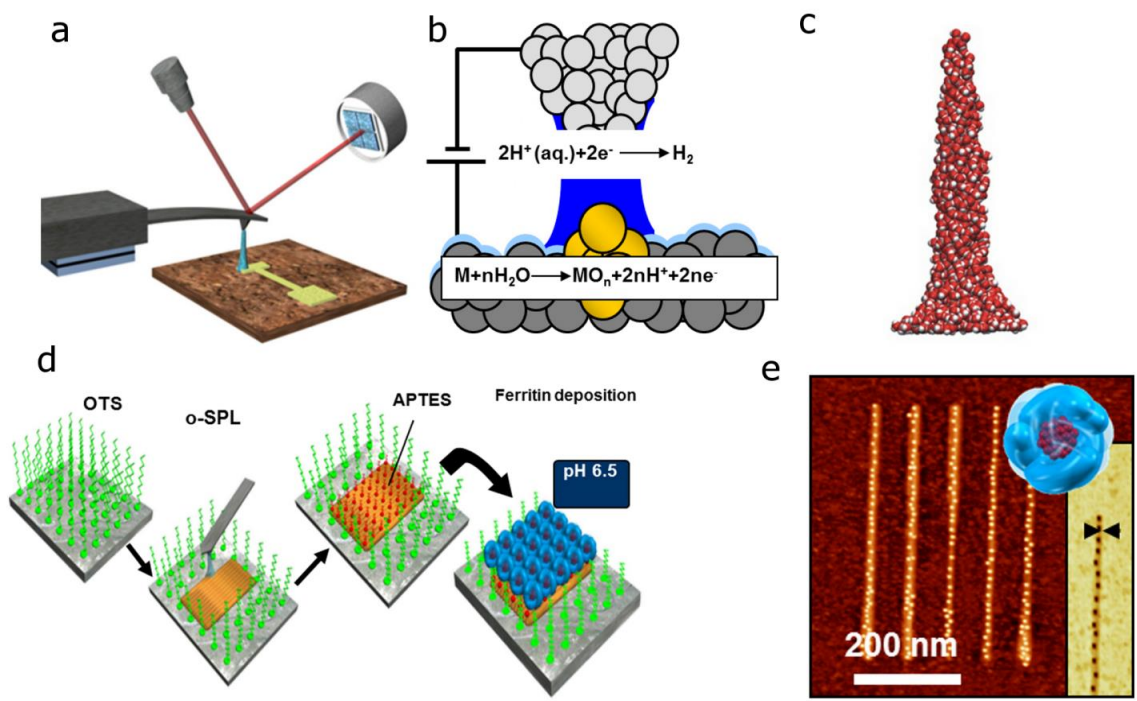

**Figure 5. Oxidation SPL**. **a**. The oxidation process used in o-SPL is mediated by the formation of water bridge which provides the oxyanions. The effective width of the liquid bridge together with the kinetics controls the feature size in o-SPL. **b**. General electrochemical reactions in local anodic oxidation. **c**. Molecular dynamics snapshot of the field-induced formation of a 2.5 nm long water bridge (1014 water molecules). Oxygen atoms are in red and hydrogen in white. **d**. Main steps to pattern ferritin proteins on a silicon surface by combining bottom-up electrostatic interactions and local oxidation. **e**. AFM image of and array of ferritin molecules. The space within the arrows is 10 nm. Panel b reprinted with permission from ref. 97. Figures reprinted with permission from: b, ref 42 © Royal Chemical Society; c, ref 97 © American Chemical Society; d,e, ref. 6, © Wiley.

The electric field has three roles in tip-based oxidation[42]. It induces the formation of the water bridge. Second, it generates the oxyanions needed for the oxidation by decomposing water molecules. Third, it drives the oxyanions to the sample interface and facilitates the oxidation process[100]. The technique generates ultra-small silicon oxide nanostructures with a lateral size between 10 and 100 nm and a height in the 1 to 10 nm range. The main parameters that control the local oxidation process are the applied voltage (from a few volts to 20-30 V), the relative humidity (20 % - 80 %), the duration of the process (10 μs – 10 s), the tip-sample distance (2 nm – 5 nm) and the scanning speed (0.5 μm/s – 1 mm/s).

**Molecular architectures.** Selective oxidation and/or complete removal of self-assembled monolayers and subsequent surface functionalization of the oxidized regions has enabled the fabrication of nanoscale architectures[6, 101]. Figure 5d-e illustrates the process to pattern linear arrays of proteins with a size that matches the molecular size of the protein, in this case ferritin. The process requires the functionalization of the silicon surface with an octadecyltrichlorosilane (OTS) monolayer, then o-SPL is applied to pattern several silicon oxide lines on the surface. The process also removes the self-assembled monolayer in the regions exposed to the field. The patterned sample is immersed in a solution containing aminopropyltriethoxysilane (APTES) molecules until an APTES monolayer is deposited in the patterned lines. Then the sample is exposed to a solution containing ferritin molecules. At a pH above 5.3 the ferritin is negatively charged. Consequently, ferritin will be attracted towards the protonated amino terminated regions of the APTES patterns. The ferritin molecules attached to the neutral OTS areas are easily removed by rinsing the sample in buffer. The complete process gives rise to the formation of arrays of proteins with a size that could match the protein size.

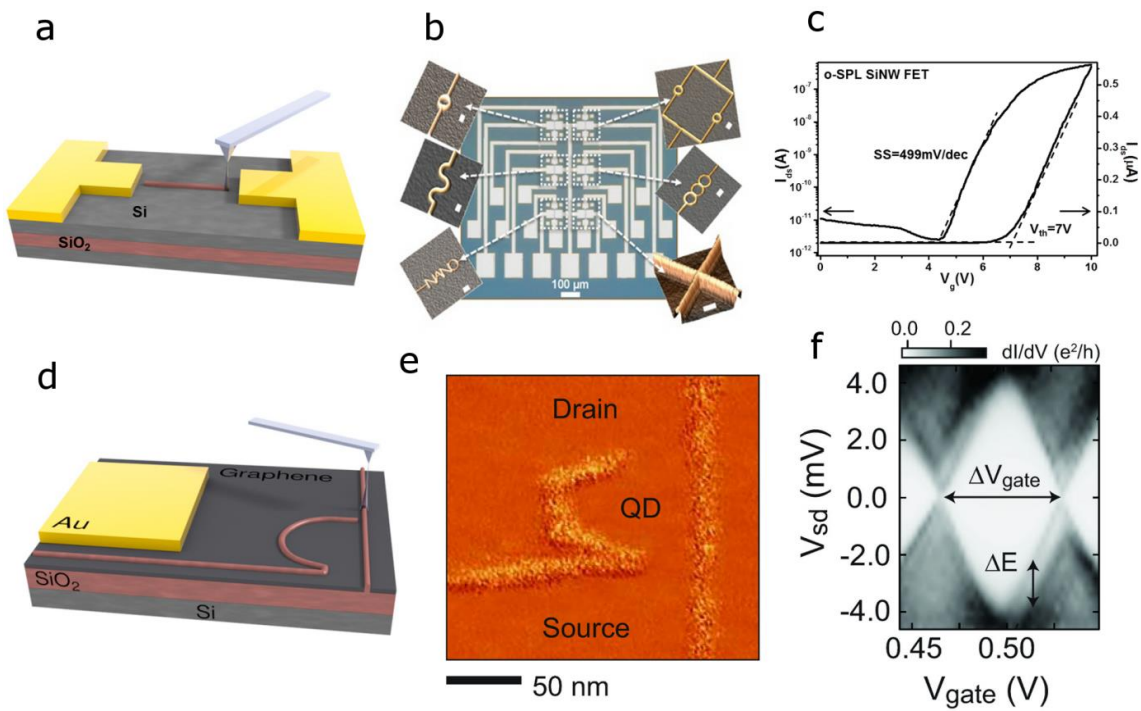

**Figure 6**. **Silicon and graphene nanolectronic devices fabricated by o-SPL**. **a**. Scheme of the fabrication of a very thin and narrow oxide mask. That mask defines the width of a silicon nanowire. **b**. AFM images of silicon nanowires of different geometries fabricated by o-SPL. The image of the gold pads and connections has been obtained by optical microscopy. Scale bar in the AFM images of 100 nm. **c**. Transfer characteristics of a silicon nanowire. **d**. Scheme of the fabrication of a graphene quantum dot. A graphene layer deposited on a silicon dioxide film is locally oxidized by an AFM tip. **e**. AFM image of a single quantum dot. **f**. Coulomb blockade diamond of the quantum dot measured at T=50 mK. Panel b reprinted with permission from ref. 15. Figures reprinted with permission from: b, ref 15, © Institute of Physics; c, ref. 102, © Institute of Physics; e,f, ef. 90 © American Institute of Physics.

**Nanoelectronic devices.** To illustrate how o-SPL is applied to fabricate nanoelectronic devices we show some of the steps into the fabrication of silicon nanowire (SiNW) transistors and graphene quantum dots. The fabrication of a silicon nanowire transistor process involves the patterning of a narrow oxide mask on top of the active layer of a silicon-on-insulator substrate[15, 61,102] (Fig. 6a). The unmasked silicon layer is then removed by using wet or dry chemical etching procedures. The local oxide protects the underneath silicon from the etching. This leaves a single-crystalline silicon nanowire with a top width that matches the width of the oxide mask. Finally, the SiNW is contacted to micrometre size platinum source and drain contacts by either photolithography or electron beam lithography. The SiNW can be transformed into a field-effect transistor by a third electrode that is usually the back of the silicon substrate. This scheme has enabled the fabrication of label-free and ultrasensitive biosensors for the detection of the early stage of recombinational DNA repair by RecA protein[103].

The fact that o-SPL does not require the use of resists that could modify the electronic properties of graphene has propitiated its application to fabricate a variety of graphene nanoscale devices. Buitelaar and co-workers[90] have fabricated a graphene-based quantum dot by locally oxidizing a graphene layer deposited on a silicon oxide surface (Fig. 6c). The microelectrodes are established by using shadow masking techniques. The quantum dot is structure is generated by locally oxidizing regions in the graphene layer. Figure 6e shows the source and drain electrodes as well as the gates and the region occupied by the quantum dot. The differential conductance as a function of the source and gate voltages shows the characteristic Coulomb diamond structure (Fig. 6f).

**Additional SPL methods**

The versatility of force microscopy to modify and manipulate surfaces (Fig. 1b) has generated some other approaches such as nanomachining[104], nanoscale dispensing[105] or dip-pen nanolithography[106]. Mechanical SPL (nanomachining) uses the mechanical force exerted by the tip to induce the selective removal of material from a surface. It has been successfully applied to modify solid substrates[107] and films of polymers[108]. The same approach has also been used for the local removal of parts of self-assembled monolayers (SAMs) and Langmuir-Blodgett (LB) films (nanoshaving)[109]. The limiting factor in creating reproducible patterns is the stability of the tip itself, which is prone to deformation and contamination from the debris of the removed material.

Nanoscale dispensing uses hollow cantilevers integrated fluidic channels to deliver small liquid drops onto a surface[110]. The fluidic channel allows soluble molecules to be dispensed through the hollow AFM tip. One remarkable application of nanoscale dispensing has been the stimulation of single living cells under physiological conditions[105].

Dip-pen scanning probe lithography (dp-SPL) offers high resolution and registration with direct write patterning capabilities[106]. The lithography functions by facilitating the direct transport of molecules to surfaces, much like the transfer of ink from a macroscopic dip-pen to paper. By depositing several different kinds of molecules on the same substrate, dp-SPL can pattern a range of desired chemistries with sub-100 nm control. Dip-pen SPL is compatible with a variety of inks, including organic and biological[111] molecules, polymers, colloidal particles, and metal ions. The intrinsic linear writing speed of dp-SPL depends on molecular transport between the probe tip and the surface, and thus it is limited by mass diffusion. The tip temperature can be used to control the ink deposition[112]. The rise of the temperature of the tip causes solid ink to melt and wet the tip. The advantage of this approach is twofold. The ink flow can be turned on and off at will whereas prior dp-SPL techniques apply ink to the surface as long as the tip stays in contact with it. In addition, the rate of the ink diffusion is tunable by controlling the tip temperature. It has been shown that poly(N-isopropylacrylamide), a type of protein adhesion molecules, can be reproducibly written from the melt[112]. These nanostructures reversibly bind and release proteins when actuated through hydrophilic-hydrophobic phase transition. This approach has also being applied to fabricate graphene nanorribons.[113]

**Large area patterning**

One of the main drawbacks of SPL techniques for technological oriented applications is the limited throughput due to the serial writing process and the required interaction time scales (Fig. 2b). Recent developments in wide area operation high speed AFM[30,114] could enhance the serial writing and imaging speed, if the pixel times are not limited by the physics of the writing process. More effectively, parallelization could enhance the throughput by a factor proportional to the number of cantilevers operating in parallel. Functional parallel probe arrays have been demonstrated in a variety of schemes ranging from fully passive systems with no actuation control and readout capability at the individual cantilever level to fully controlled systems at the individual cantilever level. For fully passive systems dp-SPL has demonstrated parallel operation of 55 k levers replicating the same nanostructures[115]. A more recent intermediate step without integrated readout employs active write control by either thermal expansion or optical addressability for polymer-pen lithography[116] and beam-pen optical lithography[117], respectively. These systems have shown the ability to pattern $cm^2$ areas, however, *in-situ* inspection is difficult and resolution is limited.

A typical single-cantilever AFM employs an optical-lever deflection scheme which cannot be easily scaled up to large cantilever arrays due to the complexity in the optical setup, signal processing, and restrictions on cantilever geometries[118]. For fully controlled parallel systems integration of actuators and sensors into the individual cantilevers is required. For a recent review of actuation and sensing schemes see ref. 119. At least three terminals are needed per cantilever to control read and write processes separately which poses an integration challenge for high numbers of parallel levers. Such a fully integrated parallel system was developed for thermo-mechanical data storage applications by IBM demonstrating parallel read and write operation at 32 nm full pitch resolution.[120] The tool used a similar cantilever design as shown in Fig. 3a solving the integration challenge by transfer of the cantilevers onto a CMOS chip. Thermal SPL and tc-SPL could directly benefit from the array technology developed for the data storage application. Array sizes up to 64x64 cantilevers have been fabricated, which would lead to a throughput increase by a factor of more than 4000. Together with the high linear speed of t-SPL throughput values of $>10^8$ µm$^2$/h are within reach which would open up new application fields such as nanoimprint master or optical mask fabrication.

Recently, parallel operation of a five-tip array for tc-SPL was demonstrated in a commercial AFM setup (see Fig. 4e)[22]. The same array is used *in situ* to pattern and image microstructures, nanowires, and complex patterns of a conjugate luminescent semiconducting polymer (see Fig. 4e), as well as conductive nanoribbons of reduced graphene oxide. Resolution down to sub-50 nm over areas of 500 µm and parallel complex 3D-patterning of conjugated polymers have been demonstrated.

Large area patterning has also been addressed by using printing-based methods. Here, the use of a stamp facilitates the up-scaling of the processes involved in some SPL methods[61,74,121-122]. In this approach, a single patterning step, say anodic oxidation in o-SPL, is replicated multiple times by using a stamp containing billions of nanostructures. The contact electrochemical replication scheme has enabled the patterning of alternating hydrophobic/hydrophilic domains on OTS monolayers[122]. However, in this approach the positioning capabilities of SPL are lost.

**Outlook**

Scanning probe lithography has experienced a quiet evolution over the last twenty years. SPL techniques offer a variety of physical and chemical approaches to modify a surface, giving rise to a wealth of methods for patterning. These methods have drawn considerable scientific attention for different reasons. On one hand, SPL is an alternative method to pattern surfaces or devices with nanoscale precision. On the other hand, SPL enables access to phenomena at the nanoscale with an easiness that is not paralleled by other techniques. Those features, together with the wide range of materials that can be patterned, the ability to pattern in ambient conditions and the relatively few requirements to transform a conventional AFM into a nanolithography instrument explain the interest and relevance of SPL in the scientific community. An illuminating example of the SPL flexibility is the recent nanofabrication of devices based on novel two-dimensional materials.

The interest and use of SPL in scientific research is established and expanding, however more technological applications still need to be fulfilled. To progress out of the research environment and into a technology used for prototyping applications, the method has to achieve sufficient throughput and reliability for day-to-day work. Some milestones towards this goal have been achieved just recently. In particular the closed-loop-lithography framework may lead to SPL tools which operate autonomously and thus with minimal learning and preparation overhead for the user. The particular strength of SPL systems of giving the user a direct feedback on the result of the operation will play a major role in the success of commercial systems. Other challenges are still to be resolved. Most prominently parallelization and tip lifetime, which although greatly enhanced for the approaches discussed in this Review, still need to be extended to meet user needs for patterning cm$^2$ areas at high speeds, high resolution and high reproducibility.

The tip lifetime defined in terms of its chemical nature and geometry is a factor that controls and determines both the reproducibility and the throughput in SPL. This Review provides an update of SPL methods based on either physical or chemical processes that better preserve the tip's geometry and chemical nature. Either because the force is exerted over polymers with stiffness orders of magnitude smaller than that one of the tip, or because the chemical process happens mostly on the sample surface.

The limited throughput of SPL methods is being addressed by two different approaches. The first approach involves the use of arrays of several SPL cantilevers, which can write and read in parallel. In SPL the actuation and topography sensing scheme can be implemented into the cantilever, which occupies an area of less than 100x100 μm$^2$. This is a unique and ideal condition for parallelization. In addition, the resolution is typically determined by the shape of the tip and thus it is not impaired by the existence of neighbouring levers. Thus in SPL parallelization the resolution is conserved while the throughput scales linearly with the number of cantilevers. The major challenges towards a highly parallel system are engineering tasks for the reliable fabrication of cantilever arrays and for a solution of the wiring problem. Both can be solved as it has been demonstrated in the "Millipede" project or in other more recent demonstrations of linear arrays of thermal cantilevers. In the near future it is expected that linear arrays of 30 thermal cantilevers can be integrated in commercial AFM, with the goal of writing and then reading more than 1 million pixels in 1 second. Since the distance between the cantilevers in the array is 100 μm, and commercial AFM scanner size is about 100 μm, 30 cantilevers will be able to nanopattern 3 mm x 0.1 mm in a single patterning action. Larger areas could be addressed by moving the sample and using fast *stitching* procedures. For example, with existing technology SPL could produce graphene nanoribbons on functionalized graphene at speeds 10$^4$ times faster than electron beam lithography. The second approach is less sophisticated, and it is based on replicating the processes involved in SPL by using micro or nano-patterned stamps.

The capability of the force microscope to provide chemical and nanomechanical information at the atomic[123], molecular and nanoscale[124, 125] levels could also be incorporated in the SPL methodology. Those methods could provide an *in-situ* metrology determination of the physical and chemical properties of the fabricated nanostructures. This is another factor that will support the evolution and expansion of SPL.

In conclusion, scanning probe lithography is approaching a stage where proof-of-principle academic experiments can become wide spread technology. The large variety of materials which can be patterned by SPL, from polymers to proteins to graphene, the high resolution, the ability to work in a range of environmental conditions, from liquid to air and vacuum, and the potential to pattern chemistry and topography simultaneously make SPL an attractive nanofabrication method for the next generation of materials and devices.

**Acknowledgements**

Financial support from the European Research Council AdG no. 340177 (RG) and StG no. 307079 (AK), the European Commission FP7-ICT-2011 no. 318804 (RG and AK), the Swiss National Science Foundation SNSF no. 200020-144464 (AK), the Ministerio de Economía y Competitividad MAT2013-44858-R (RG), the National Science Foundation CMMI-1100290 (ER), the MRSEC program DMR-0820382 (ER), and the Office of Basic Energy Sciences of the Department of Energy DE-SC0002245 (ER) are acknowledged.